# BIM can help decarbonize the construction sector: life cycle evidence from Pavement Management Systems

Anne de Bortoli[1,2,3*], Yacine Baouch[4], Mustapha Masdan[2]

**1** CIRAIG, École Polytechnique de Montréal, P.O. Box 6079, Montréal, Québec, H3C 3A7, Canada

**2** Direction technique, Eurovia Management, 18 Place de l'Europe, 92500, Rueil-Malmaison, France

**3** LVMT, Ecole des Ponts ParisTech, Cité Descartes, 6-8 Avenue Blaise Pascal, 77420 Champs-sur-Marne, France

**4** Université Technologique de Compiègne, Recherche de Royallieu, rue du Docteur Schweitzer, 60203 Compiègne, France

**\*** Corresponding author; e-mail: anne.debortoli@polymtl.ca

ABSTRACT

Transforming the construction sector is key to reaching net-zero, and many stakeholders expect its decarbonization through digitalization, but no quantified evidence has been brought to date. This article proposes the first environmental quantification of the impact of Building Information Modeling (BIM) in the construction sector. Specifically, the direct and indirect greenhouse gas (GHG) emissions generated by a monofunctional BIM to plan road maintenance – a Pavement Management System (PMS) - are evaluated using field data from France. The related carbon footprints are calculated following a life cycle approach, using different sources of data – including ecoinvent v3.6 – and the IPCC 2013 GWP 100a characterization factors. Three design-build-maintain pavement alternatives are compared: scenario 1 relates to a massive design and surface maintenance, scenario 2 to a progressive design and pre-planned structural maintenance, and scenario 3 to a progressive design and tailored structural maintenance supported by the PMS. First, results show the negligible direct emissions due to the PMS existence – 0.02% of the life cycle emissions of scenario 3's pavement, e.g. 0.52 t CO2eq for 10 km and 30 years. Second, the base case and two complementary sensitivity analyses show that the use of a PMS is climate-positive over the life cycle when pavement subgrade bearing capacity improves over time, neutral for the climate otherwise. The GHG emissions savings using BIM can reach up to 14 and 30% of the life cycle emissions respectively compared to scenario 2 and 1, and resp. 47 and 65% when restraining the scope to maintenance and rehabilitation and excluding original pavement construction. Third, the neutral effect of BIM in case of a deterioration of the bearing capacity of the subgrade may be explained by design practices and safety margins, that could in fact be enhanced using BIM. Fourth, the decarbonization potential of a multifunctional BIM is discussed, and research perspectives are presented.

# 1  Introduction and background

Construction is a key sector that must be transformed to reach a net-zero society, and many stakeholders expect its decarbonization through digitalization, but no quantified evidence has been shown to date. In Canada, construction is the second most carbon-intensive sector, accounting for 12% of the national emissions, 59% of them coming from infrastructure (de Bortoli and Agez, Under review). To decrease this burden, many green construction practices have been appraised using Life Cycle Assessment (LCA) - for buildings (e.g. Anand and Amor, 2017; Vilches et al., 2017) or infrastructure (e.g. AzariJafari et al., 2016; Saxe et al., 2020) - while the environmental consequences of digitalization such as the use of Building Information Modelling (BIM) for buildings have been presented positively as a way to automatize environmental quantifications and optimizations (Soust-Verdaguer et al., 2017). BIM is defined as the "use of a shared digital representation of a built asset to facilitate design, construction, and operation processes to form a reliable basis for decisions" (International Organization for Standardization, 2018). Its direct and indirect environmental impacts – those due to resp. the amortization of BIM's hardware, equipment and infrastructure, their operation, and the usage of BIM's software, and the consequences of BIM on construction processes and management - have never been quantified to date. In the meantime, the environmental impact of digital services has been increasingly studied (ADEME, 2016) and shown to be significant:

Information and Communication Technologies (ICT) would emit between 1.8 and 3.9% of the global anthropogenic greenhouse gas (GHG) (Freitag et al., 2021). Thus, it cannot be inferred that the direct environmental impact of BIM is negligible, nor that its potential indirect environmental benefits overcome its direct costs. This paper aims to offer a first life cycle quantification of these aspects.

The impact of infrastructure construction exceeding those of buildings (de Bortoli and Agez, Under review), this paper will focus on infrastructure. Moreover, among the different kinds of infrastructure, roads support the most emitting mean of transportation (Our world in data, 2020). Their decarbonization is thus a priority to reach a net-zero pathway. However, BIM for infrastructure is less advanced than BIM for buildings (Malagnino et al., 2021). "I-BIM", "Infra-BIM", or BIM for infrastructure, started to be implemented in 2013 in the rail construction sector of developed countries (Matejov and Šestáková, 2021). Since then, it has been shown to help design road elements thanks to visualization – e.g. pavement sections, roundabouts, tunnels (Vignali et al., 2021) -, planning airport maintenance (Abbondati et al., 2020), and in general improving infrastructure (Costin et al., 2018). I-BIM-based environmental assessments would also increase sustainability awareness within design teams (van Eldik et al., 2020), but the direct and indirect environmental impacts of I-BIM have not been quantitatively investigated. The difficulty in assessing these impacts lies both in the versatility of I-BIM and its novelty. It started to be punctually used one decade ago, and 47 categories of usages have been classified, including automations, pricing, monitoring, failure detection, and maintenance planning (Costin et al., 2018).

Pavement Management Systems (PMS) are common software programs analyzing road condition data to plan maintenance. These PMSs allow to adopt different strategies in the design-build-maintain sequence related to a pavement life cycle, compared to a non-digitally helped pavement management. They are thus equivalent to a monofunctional I-BIM dedicated to road maintenance and used for decades by road operators. As ex-post assessments generally present a lower degree of uncertainty than prospective assessments, a case study will be conducted to quantify on field data the consequence of using this experienced BIM function on a pavement life cycle carbon footprint, compared to alternative ways to design and maintain this pavement.

Two main schools of thought exist to design roads and schedule their maintenance: empirical vs mechanical-empirical (ME) methods. The American continent mainly used the AASHTO empirical method (AASHTO, 1993), before switching progressively to a ME version (AASHTO, 2008; US DoT - FHWA, 2019a, 2019b, 2019c, 2019d). In France and Africa, the LCPC-Setra ME method is widely used (LCPC-Sétra, 1994). A catalogue of standard solutions has been made available (Corte et al., 1998), while design and maintenance plans of strategic pavements are usually tailored using the match ME software "Alize". As they carry a substantial portion of the traffic (around 20%) on a short portion of the national road network (1%) (de Bortoli 2018), high-traffic roads present strategic socioeconomic importance, and more design-build-maintain alternatives. Thus, this case study will focus on high-traffic roads. The objectives of this study are to quantify the carbon footprint of (1) the PMS function of an I-BIM based on field data, and (2) three design-build-maintain alternatives for a high-traffic road in France over its entire life cycle, (3) to understand the potential impact of the PMS function of a BIM on climate change, and (4) discuss the environmental consequences of a multifunctional BIM.

# 2 Method

## *2.1 Overview of the method*

A method is developed to quantify the carbon footprint of the most common ways to design, build and maintain high-traffic roads in France, including an option where maintenance is digitally helped using a PMS, to compare the GHG emissions related to these different practices and seize the consequences of using digital tools for maintenance operation planning that could be included in an I-BIM. This method consists in developing three design-build-maintenance scenarios to compare their life cycle carbon footprint, calculated based on foreground and background inventories (or emission factors) from the literature, road maintenance operators, and their suppliers. The quantification method is based on LCA, applied consistently with Standards ISO 14040 and 14044 (International Organization for Standardization, 2006a, 2006b), and the characterization factors chosen are IPCC 2013 GWP 100a, the most recent factors when this study was conducted (mid-2021). The system boundaries include production, maintenance, and use stages. The functional unit is "building and maintaining in good condition over 30 years a 10 kilometer-long and 7-meter-wide section of highway in France, under a traffic of 500 heavy vehicles per day".

In this method section, the foreground modeling will be first presented, i.e. the scenarios of design-build-maintenance operations and related PMS activities as well as the equations to calculate the carbon footprint of the pavement over its life for each scenario. The scenarios are detailed in the following section developed with pavement construction companies and highway concessionaires in France. Second, the background data and emission factors (EF) to quantify this carbon footprint will be detailed. An EF is the carbon footprint of any unitary activity occurring during the pavement lifespan.

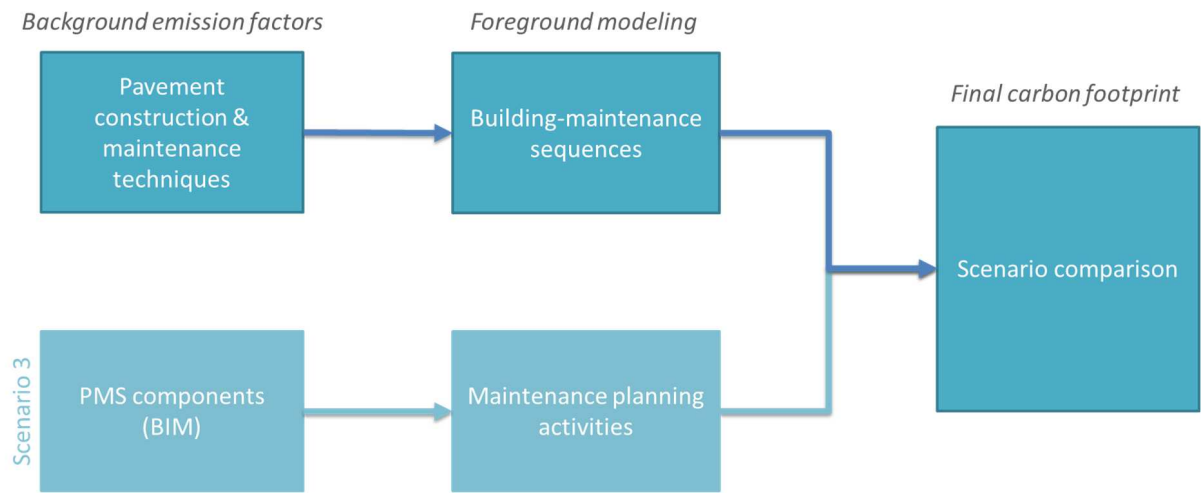

**Figure 1** Calculation method overview

## 2.2 Case study scenarios

### 2.2.1 A theoretical representative French highway

The case study is conducted on a theoretical representative French highway, with two lanes in each direction, bearing a traffic of 500 heavy vehicles per day and direction (the "T1" French class of Average Annual Daily Traffic (AADT)), with a traffic rate assumed at 0%. A 10-kilometer-long section designed to last 30 years is considered, with a width of 7 meters per direction (excluding shoulders), and a class « PF2qs » subgrade at the commissioning stage. Subgrades are classified in France based on the French standard NF P 94-117-1 (AFNOR, 2000), the "SETRA-LCPC" guides on pavement subgrade design (Corte et al., 2000a, 2000b), and a national complement on PF2qs class subgrades (CEREMA, 2017). A PF2qs subgrade corresponds to a "Plate Test Static Deformation Module" (EV2) between 80 MPa (included) and 120 MPa (excluded) (CEREMA, 2017). Ten years after commissioning, this bearing capacity is assumed to decrease toward a "PF2" subgrade class (50 MPa $\leq$ EV2 < 80 MPa) over 25% of the linear and increase toward a "PF3" subgrade (120 MPa $\leq$ EV2 < 200 MPa) over 25% of the section as well, respectively due to drainage issues deteriorating the subgrade and traffic compacting the soil. Fifty percent of the linear will maintain a PF2qs subgrade overtime.

This evolution has been arbitrarily estimated by Eurovia, a leading road construction company worldwide, and sensitivity analyses will be conducted, respectively over a restricted and an extreme range of subgrade evolutions. This problem presents three parameters – respectively %PF2, %PF2qs and %PF3 the average percentage of each subgrade class length over the pavement lifespan. The sum of these parameters equals 100% but can vary independently. Thus, there is no continuous representation showing all the possibilities of the subgrade evolutions. For this reason, arbitrary values are chosen for these sensitivity analyses. Table 1 presents the parameters' values that will be tested for the restricted range sensitivity analysis (SA), while Table 2 presents the parameters' values for the extreme range SA.

**Table 1 Parameters' values for the "restricted range" sensitivity analysis**

|        | a   | b   | c   | d   | e   | f - base | g   | h   | i   | j   | k   |
|--------|-----|-----|-----|-----|-----|----------|-----|-----|-----|-----|-----|
| %PF2   | 0.5 | 0.4 | 0.3 | 0.2 | 0.1 | 0        | 0   | 0   | 0   | 0   | 0   |
| %PF2qs | 0.5 | 0.6 | 0.7 | 0.8 | 0.9 | 1        | 0.9 | 0.8 | 0.7 | 0.6 | 0.5 |
| %PF3   | 0   | 0   | 0   | 0   | 0   | 0        | 0.1 | 0.2 | 0.3 | 0.4 | 0.5 |

**Table 2 Parameters' values for the "extreme range" sensitivity analysis**

|        | a | b   | c   | d   | e   | f - base | g   | h   | i   | j   | k |
|--------|---|-----|-----|-----|-----|----------|-----|-----|-----|-----|---|
| %PF2   | 1 | 0.8 | 0.5 | 0.5 | 0.4 | 0.25     | 0.1 | 0.1 | 0   | 0   | 0 |
| %PF2qs | 0 | 0.2 | 0.5 | 0.4 | 0.5 | 0.5      | 0.5 | 0.4 | 0.5 | 0.2 | 0 |
| %PF3   | 0 | 0   | 0   | 0.1 | 0.1 | 0.25     | 0.4 | 0.5 | 0.5 | 0.8 | 1 |

*2.2.2 Massive vs progressive designs and maintenance*

Two ways of designing pavements can be considered in France: a massive vs a progressive design. We call massive design (Scenario 1) an approach that consists of designing a pavement to last till the end of its life without structural reinforcement under the traffic expected. In that case, maintenance operations are only performed to keep the pavement surface in good condition: ensuring waterproofness, a skid resistance suiting safety thresholds, and good riding conditions. In France, massive design is guided by the LCPC-Setra catalogue for pavements (Corte et al., 1998).

Alternatively, a progressive design (Scenarios 2 and 3) is performed when a pavement is not originally designed to mechanically resist the traffic expected over its entire service life and needs structurally-reinforcing maintenance operations that will at the same time restore good surface condition. This second approach requires to use a ME design software and may be selected by road concessionaires to optimize discounted cash flows and reduce traffic risk. This risk management practice is particularly attractive under an ever more uncertain future, due to climate change consequences on pavement conditions and changes in behaviors related to cultural standards evolutions, economic crises, and else.

Then, two approaches allow for managing the maintenance of progressively-designed pavements: theoretical mechanical planning (scenario 2) or data-supported adaptative planning (scenario 3). The maintenance in scenario 2 is planned during the design process using a ME design software considering traffic forecast and can be further modified with traffic data. Scenario 3 consists of data-supported maintenance management: it relies on traffic data but also pavement condition data collected and monitored over time, stored, and analyzed in a PMS. This is the most tailored maintenance approach existing, as it adapts to the real evolution of traffic, climate, and thus pavement mechanical conditions. It thus avoids premature pavement failures as well as structural oversizing.

*2.2.3  Design and maintenance sequence by scenario*

**Description**

The pavement design and maintenance sequences of the three scenarios are illustrated in Figure 2, with a theoretical risk of failure of 5%. It means that, over the pavement's lifespan, with the original subgrade class and under the traffic expected, 5% of the surface will have experienced bottom-up cracking calling for rehabilitation (=full-thickness reconstruction). The thickness and material type of each layer are presented over the 30-year service life in Figure 2. The different materials and techniques used are cold micro asphalt concrete surfacing (CMACS, called "enrobés coulés à froid" (ECF) in French), semi-coarse aggregate asphalt concrete (overlays) (SCAC(O), called "béton bitumineux semi-grenu" (BBSG) in French), Road base asphalt (RBA, called "grave-bitume" (GB3) in French), (very) thin asphalt concrete overlays ((V)TACO, called "béton bitumineux (très) mince (BB(T)M) in French). Scenario 1 is extracted from the French LCPC-Setra Catalogue (Corte et al., 1998). SCAC M&F corresponds to mill (M) a former surface layer before filling (F) it with SCAC. Scenario 2 is designed with ODIN, Eurovia's ME design software based on the same physics as the standard Alizé's software. In Scenarios 1 and 2, the road manager does not have data on the evolution of the road condition and is thus unaware of the evolution of the subgrade bearing capacity. This results in subsections of the pavement failing prematurely. The cumulated section's portion likely to fail over time is calculated with Odin and provided in the supplementary material. Finally, scenario 3 is also managed with Odin but with perfect knowledge of the pavement evolution, including the subgrade's modulus, and maintenance is optimized accordingly.

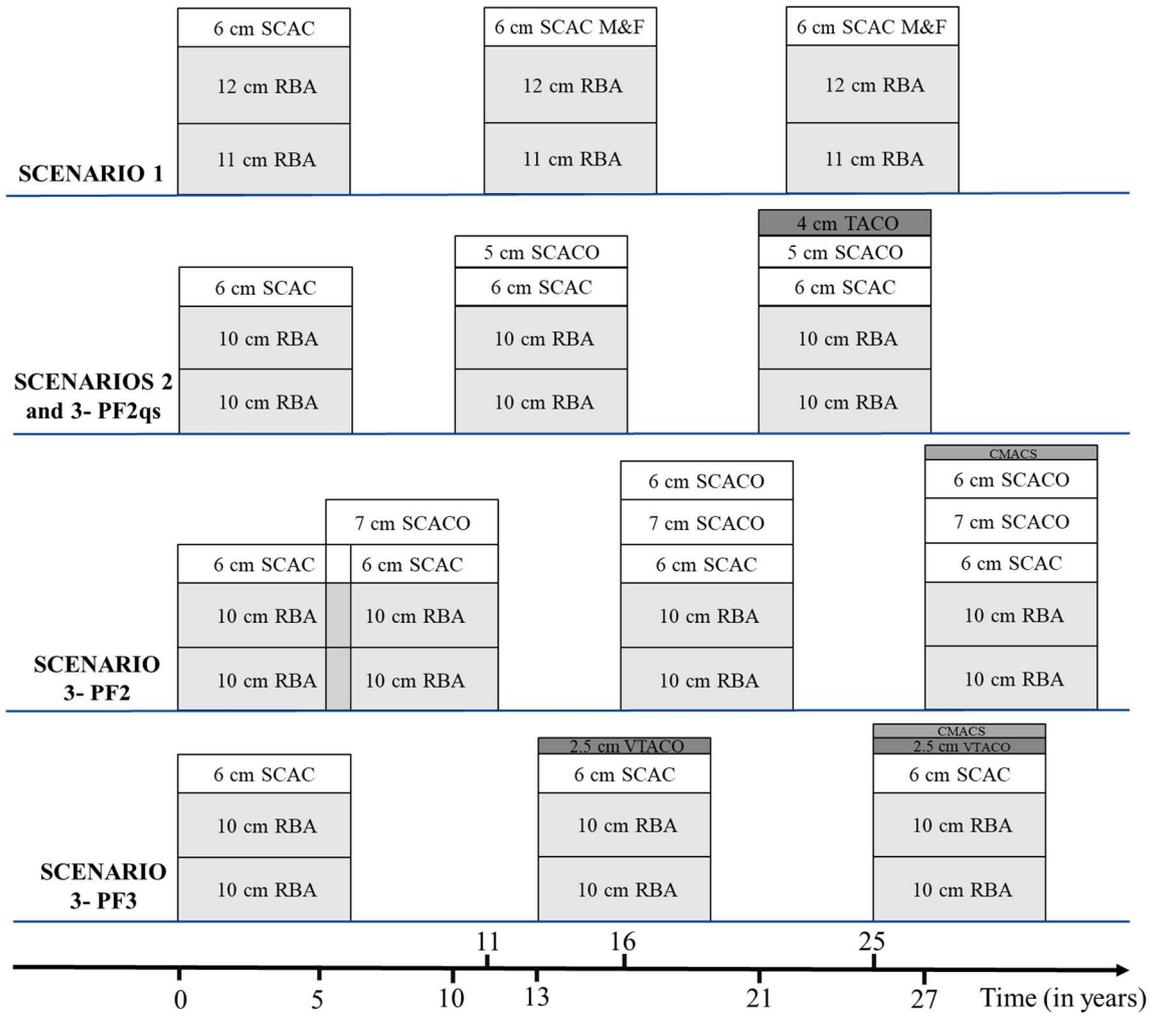

**Figure 2 Design and maintenance sequence of the different scenarios**

**Carbon footprint**

The carbon footprint $CF$ of the built-maintained pavements over 30 years will be calculated following equation (1), where $W_{section}$ is the width of the pavement studied (in meters), $L_{section}$ the length (in meters), and $EF_i$ the emission factor of the construction operation $i$ occurring during the pavement lifespan (in kgCO$_2$eq/sm), this operation being related to original construction or maintenance.

$$CF(built-maintained\ pavement) = \sum_i W_{section} \times L_{section} \times EF_i \qquad (1)$$

*2.2.4   Pavement Management*

**Modeling scenarios**

In scenarios 1 and 2, premature pavement failures are identified by regular patrolling, i.e. via visual surveying performed by trained staff, often from the pavement management company. As scenario 3 also comes with this patrolling, these operations will be removed from the system boundaries for all scenarios, following ISO 14044 guidelines (International Organization for Standardization, 2006b). On the other hand, scenario 3 requires pavement monitoring by an equipped truck which measures the evolution of the bearing capacity of the highway pavement. This data is therefore collected and then stored in a database. A pavement asset management team will then analyze them to plan the maintenance operations.

To model this PMS's features (illustrated in Pavement Management System use casesFigure 3), a "SysML Use cases" diagram is used. this formalism describes use cases and external actors' interactions in a concise way. Tree external system users are considered: external data (that need to be collected), an external database (that is used to store data), and the user of the PMS (that is a member of the pavement asset management team). Four use cases are considered: collecting external data, storing internal data, running a local task, and transferring internal data. The last two use cases constitute the PMS's maintenance operation planning. To be more efficient, only the main use cases are represented, and the extend/include relationships are omitted. For example, collecting data may be extended to storing.

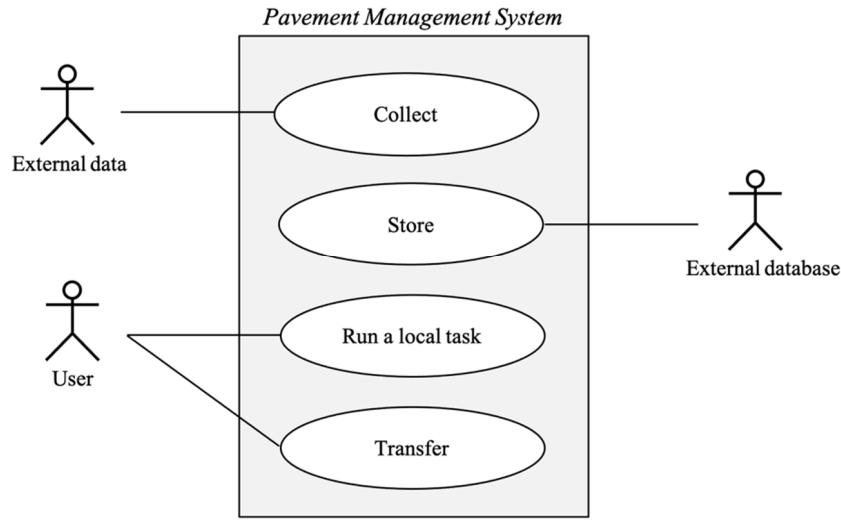

Figure 3 Pavement Management System use cases

**PMS carbon footprint**

The Global Warming Potential (GWP) of a PMS for a pavement section, $I_{PMS}^{section}$, can be calculated using the generic equation (2):

$$I_{PMS}^{section} = I_{PMS}^{data} + I_{PMS}^{storage} + I_{PMS}^{maintenance} \qquad (2)$$

Where:

- The section is defined by a length and a maintenance period

- $I_{PMS}^{data}$ is the GWP to collect the data related to the section

- $I_{PMS}^{storage}$ is the GWP to store the data related to the section

- $I_{PMS}^{maintenance}$ is the GWP to plan the maintenance operation related to the section

**Data collection of the pavement condition**

In France, we use different types of deflectometer to measure pavement deflection. In this case study, a curviameter is used, i.e. an instrumented truck weighted with 13 tons on its rear axle to record the deformation under this axle. According to ecoinvent's truck classification, this vehicle is considered as a 16-32 ton truck, and a EURO6 emissions standard is considered.

French national roads are standardly monitored every three years in each direction (Lorino et al., 2006), but some strategic roads can be monitored more often, such as once a year. The GWP of the data collection, $I_{PMS}^{data}$, is defined by equation (3):

$$I_{PMS}^{data} = n_{round} * P_m * L_{section} * F_m \qquad (3)$$

Where:

- $P_m$ is the maintenance period in years
- $L_{section}$ is the length of the studied section in km
- $n_{round}$ is the number of monitoring rounds per year
- $F_m$ is the emission factor of the vehicle used to collect road condition data in kg $CO_2$eq per km

We choose the worst-case scenario, in which the monitoring round is conducted every year. The emission factor of the instrumented truck, $F_m$, is calculated from the 16-32 ton EURO6 Transport freight process from Ecoinvent 3.4 process (see Table 4). To estimate the carbon footprint of such a truck per kilometer traveled, we multiply the freight transportation process based on the ton-kilometer functional unit by the average load considered by ecoinvent – i.e. 5.79 tons (Spielmann et al., 2007).

**Data storage**

The PMS's database is composed of several pieces of hardware: 3 servers (ProLiant BL460c Gen10 and ProLiant BL460c Gen9 models), 1 storage bay (3PAR 8200 model), and 1 backup bay (StorOnce 5500 model). The GWP of the data storage related to the studied section, $I_{PMS}^{storage}$, is defined by equation (4):

$$I_{PMS}^{storage} = \frac{P_m * L_{section}}{L_t} * k_{database} * \left(\frac{I_{hard}}{\Delta_{hard}} + I_{yrun}\right) \qquad (4)$$

Where:

- $L_t$ is the total section's length informed in the database in km

- $k_{database}$ is the allocation coefficient considering the proportion of the database used for the PMS
- $I_{hard}$ is the GWP of the PMS's database hardware in kg CO2eq
- $\Delta_{hard}$ is the depreciation period of the hardware database in years
- $I_{yrun}$ is the GWP of a one-year run of the database in kg CO2eq

In this case, the entire database is dedicated to PMS: $k_{database} = 1$. $\Delta_{hard}$ and $L_t$ are given by an internal expert (see Table 3). $I_{hard}$ and $I_{yrun}$ are provided by the manufacturer (see Table 4).

**Data analyses**

Data are analyzed by the pavement asset management team to plan the maintenance. The evolution of the bearing capacity of the subgrade is extrapolated from the pavement deflection. Then, the need for further structural reinforcements or delaying operations is estimated. The digital activities of the team are composed of local tasks, with a computer, and data transfer from the PMS's database. More specifically, two types of users, and thus computers, are involved: standard and advanced stations. The GWP of the maintenance operation planning, $I_{PMS}^{maintenance}$, is defined by equation (5):

$$I_{PMS}^{maintenance} = n_{operation} * P_m * L_{section} * (k_t * F_{transfer} + F_{adv} + F_s) \qquad (5)$$

Where:

- $n_{operation}$ is the number of maintenance planning during one year
- $k_t$ is the data volume transferred per PMS planning operation per km and year
- $F_{transfer}$ is the emission factor to transfer data
- $F_s$ and $F_{adv}$ are the emission factors of the amortization of the two types of computing stations

According to an internal expert, maintenance operation planning is carried out every 15 years: $n_{operation} = 1/15$. This internal expert considers, in a worst-case scenario, that three maintenance planning conducted on a 100 km section over 50 years require 50 gigabytes (GB) of data transfer. Thus, $k_t$ is calculated from this estimation.

$F_{adv}$ and $F_s$ are calculated with equation (6):

$$F_{adv} = k_{adv} * \left( \frac{I_{adv\_device}}{\Delta_{adv\_device}} + 24 * P_{adv\_device} * F_{elec} \right) \quad (6)$$

Where:

- $k_{adv}$ is the number of days spent on the PMS software on the advanced station, per PMS planning operation, per km, and per year
- $I_{adv\_device}$ is the GWP of the advanced computing station's manufacturing in kg CO2eq
- $\Delta_{adv\_device}$ is the depreciation period of the advanced computing station
- 24 is the number of hours per a day
- $P_{adv\_device}$ is the electrical power in watts, daily weighted
- $F_{elec}$ is the emission factor of the electricity consumption

The internal expert considers, in a worst-case scenario, that three maintenance operation planning conducted on a 100 km-long section for 50 years require 40 days on the PMS software on the standard station and 15 days on the PMS software on the advanced station. This allows calculating $k_{adv}$ and $k_s$ (see Table 3). A standard computer is composed of a laptop and two display screens while an advanced computer uses a computer desktop with two display screens, a pointing device, and a keyboard. This allows calculating $I_{adv\_device}$ and $I_{s\_device}$ from the ecoinvent database (see Table 3). $\Delta_{adv\_device}$ and $\Delta_{s\_device}$ correspond to 5 years, converted to 1825 days.

We assume, in agreement with the internal expert, that a display screen operates for 8 hours per day at 90 W and stays on stand-by at 1 W for the remaining time. A laptop also operates 8 hours per day at 65 W but is turned off the rest of the time. Finally, a desktop computer operates all day, i.e. 24 hours, at 300 W. Those assumptions allow us to determine $P_{adv\_device}$ and $P_{s\_device}$ (see Table 3). The emission factor of the electricity consumption $F_{elec}$ is also given (see Table 4).

**Table 3 Foreground data of PMS carbon footprint**

| Data | Value | Unit |
|---|---|---|
| $n_{round}$ | 1 | |
| $k_{database}$ | 1 | |
| $\Delta_{hard}$ | 5 | years |
| $L_t$ | 2737 | km |
| $n_{operation}$ | 1/15 | |
| $k_t$ | $3.33*10^{-3}$ | GB*(operation * year * km)$^{-1}$ |
| $\Delta_{adv\_device}$ | 1825 | days |
| $\Delta_{s\_device}$ | 1825 | days |
| $k_{adv}$ | $10^{-3}$ | (operation * year * km)$^{-1}$ |
| $k_s$ | $2.6*10^{-3}$ | (operation * year * km)$^{-1}$ |
| $P_{adv\_device}$ | 330.67 | W |
| $P_{s\_device}$ | 83 | W |

## 2.3  Background emission factors

### 2.3.1  Road construction and maintenance operations

Life cycle inventories (LCI) for road construction and maintenance operations have been developed by de Bortoli et al. (2018; 2022a) in the French context. They are the most representative LCIs within the existing LCIs for the context of this study (de Bortoli, 2020) in terms of asphalt binder, aggregates, asphalt mixing, building machines, material transportation, and else. These cradle-to-laid LCIs are based on the functional unit of one square meter of pavement built over a certain thickness, in centimeters, or milled. As these LCIs were specifically developed to study resurfacing strategies, some of them have been re-scaled to model the original construction of the pavement, over a thicker layer, consisting of more material per square meter but also sometimes more compaction operations. The compaction of the subbase and base layers has been modeled based on Eurovia practices depending on the layer's thickness. The detail can be found in the supplementary material (excel spreadsheet).

*2.3.2 Pavement management system*

**Data storage**

$I_{hard}$ and $I_{yrun}$ are provided by the manufacturer to the internal expert (see Table 4).

**Data analyses**

$F_{elec}$ is the emission factor of the electricity mix. The electricity is mainly consumed in France and we use the French *Base carbone* emission factor (ADEME, n.d.). $F_{transfer}$ is calculated based on the worst energy intensity of Internet transmission reported by Coroama and Hilty, i.e. 1.8 kWh/GB (2014), and $F_{elec}$. To calculate $I_{adv\_device}$ and $I_{s\_device}$, we use the following global market ({GLO}) processes from ecoinvent v3.4: *"Computer laptop"* and *"Display liquid crystal 17 inches"* for the standard station, and *"Computer, desktop, without screen"*, *"Keyboard"* and *"Pointing device optical mouse with cable"* for the advanced station.

**Table 4 Background data to calculate the PMS's carbon footprint**

| Parameter | Value | Unit |
|---|---|---|
| $F_m$ | 0.92 | kg CO2eq * km$^{-1}$ |
| $I_{hard}$ | 167 | kg CO2eq |
| $I_{yrun}$ | 1165 | kg CO2eq * year$^{-1}$ |
| $F_{elec}$ | 59.9*10$^{-3}$ | kg CO2eq * kWh$^{-1}$ |
| $F_{transfert}$ | 107.82*10$^{-3}$ | kg CO2eq * GB$^{-1}$ |
| $I_{adv\_device}$ | 1052.43 | kg CO2eq |
| $I_{s\_device}$ | 942.31 | kg CO2eq |

# 3 Results and interpretation

## *3.1 Comparison of the three design-build-maintain practices*

Figure 4 shows the carbon footprint of the pavement over its 30-year-long lifespan depending on the scenario. Scenarios 2 and 3 emit less than scenario 1, respectively by 19 and 22%. This means savings of respectively 961 and 1076 t $CO_2$eq for the 10 km-long section over 30 years. Thus, massive GHG savings occur from the transition from massive to progressive pavement design. But tailored maintenance thanks to the PMS in scenario 3 reduces by an additional 3% the GHG emissions of the pavement over its life cycle compared to a non-data-driven maintenance scheme (scenario 2). If not considering the original construction emissions, BIM generates 11% GHG savings on the maintenance and rehabilitation of the pavement over its service life. The direct impact of the PMS usage – e.g. manufacturing of the digital hardware, infrastructure and data collection vehicle as well as their operation and the use of the software - is highlighted in red (Figure 4): it accounts for 0.52 t $CO_2$eq over the pavement section life cycle, e.g. 0.02% of scenario 3's carbon footprint, for one data collection per year. If the pavement condition data is collected every three years instead, the PMS direct impact is even lower, decreasing to 0.33 t $CO_2$eq. The carbon footprint of the PMS is negligible compared to the pavement construction and maintenance emissions. On the contrary, it might be surprising to get more emission gains limited to 3.4% from the tailored maintenance.

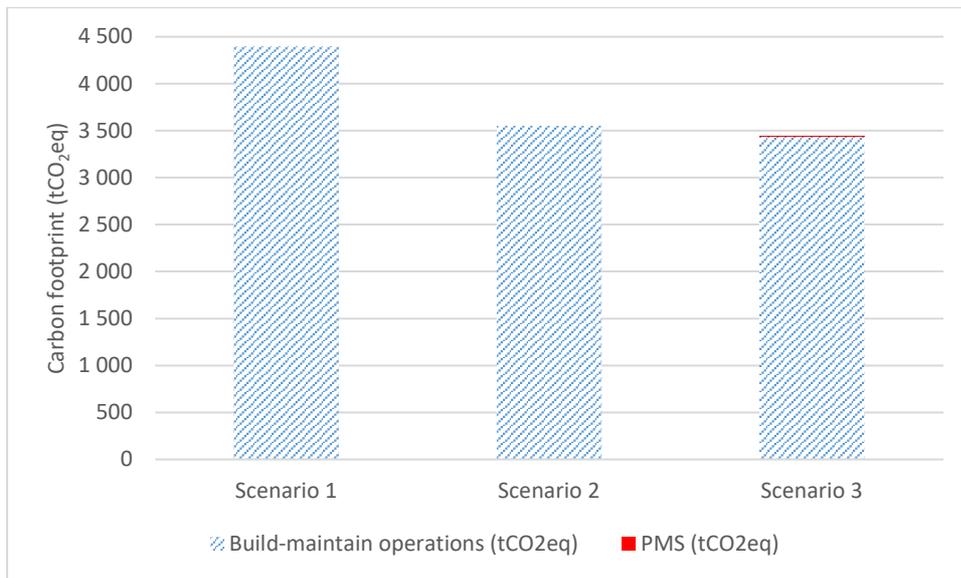

**Figure 4 Carbon footprint of the pavement life cycle depending on the scenario**

### *3.2 Roadworks' contributions*

To understand better why the PMS does not bring more substantial gains (-3.3% emissions), Figure 5 presents the impact contribution of each stage of the pavement life cycle. The use of the PMS allows to reduce the probability of pavement damage and thus the length to rebuild: scenario 2 presents a reconstruction stage emitting 2.37 kg $CO_2$eq/sm when it represents 2.74 kg $CO_2$eq/sm for scenario 3. The effects of the PMS on maintenance operation gains are a bit more important: the maintenance stage emits 13.3 kg $CO_2$eq/sm in the case of scenario 2, against 11.3 kg CO2eq/sm for scenario 3.

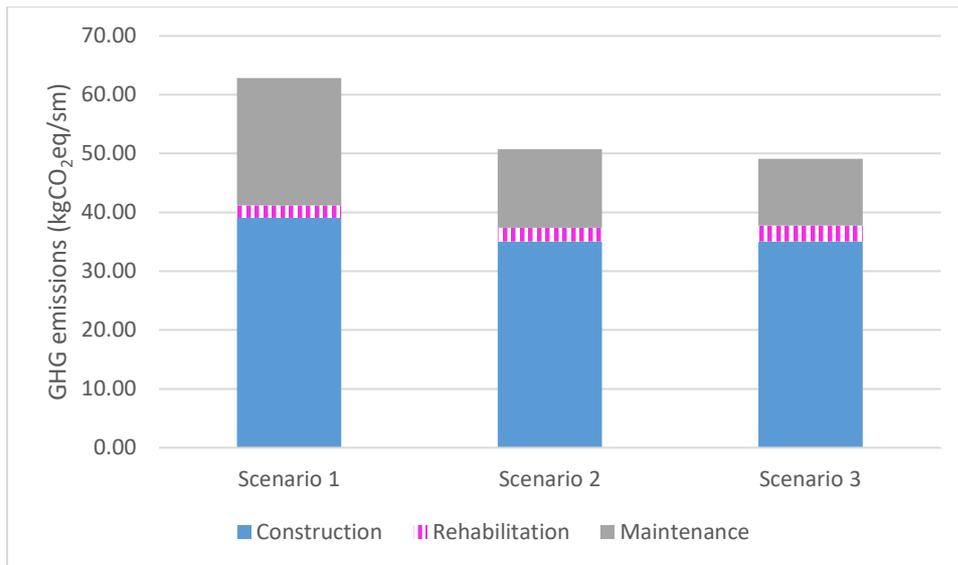

**Figure 5 GHG emissions of the pavement life cycle for each stage of its life cycle, per square meter**

### *3.3 PMS contributions*

All results, presented in Table 5 and Figure 6, relate to the 10 km-long section over 30 years. The GWP of the PMS is 520.7 kg $CO_2$eq. It is divided in 241.1 kg $CO_2$eq (46.3%) for data storage, 277.9 kg $CO_2$eq (53.4%) for data collection activities and 1.7 kg $CO_2$eq (0.3%) for maintenance planning activities, e.g. the usage of the PMS by asset managers. Data collection is the biggest contributor to PMS's carbon footprint, with 53.4% of the total impact. Data storage is almost equally contributing, with 46.3% of the total GWP, while the PMS's usage itself is responsible for a negligible amount of GHGs (0.3%). The GWP due to data storage is largely due to the database running, i.e. electricity consumption (97.3%). The amortization of the database's hardware manufacturing represents only 2.7% of the total GWP. The GWP of maintenance planning activities comes from the local tasks (87.5%) and data transfer tasks (12.5%).

**Table 5 PMS's carbon footprint and contributions to total**

|  | *GWP (kg $CO_2$eq)* |
|---|---|
| PMS total | 520.8 |

| | | | |
|---|---|---|---|
| Data storage | 241.1 | 46.3 % | |
| *Of which: hardware* | *6,5* | | *2.7 %* |
| *Of which: running* | *234.5* | | *97.3 %* |
| Data collection | 277.9 | 53.4 % | |
| Maintenance planning | 1.7 | 0.3 % | |
| *Of which: local task* | *1.5* | | *87.6 %* |
| *Of which: transfer task* | *.2* | | *12.4 %* |

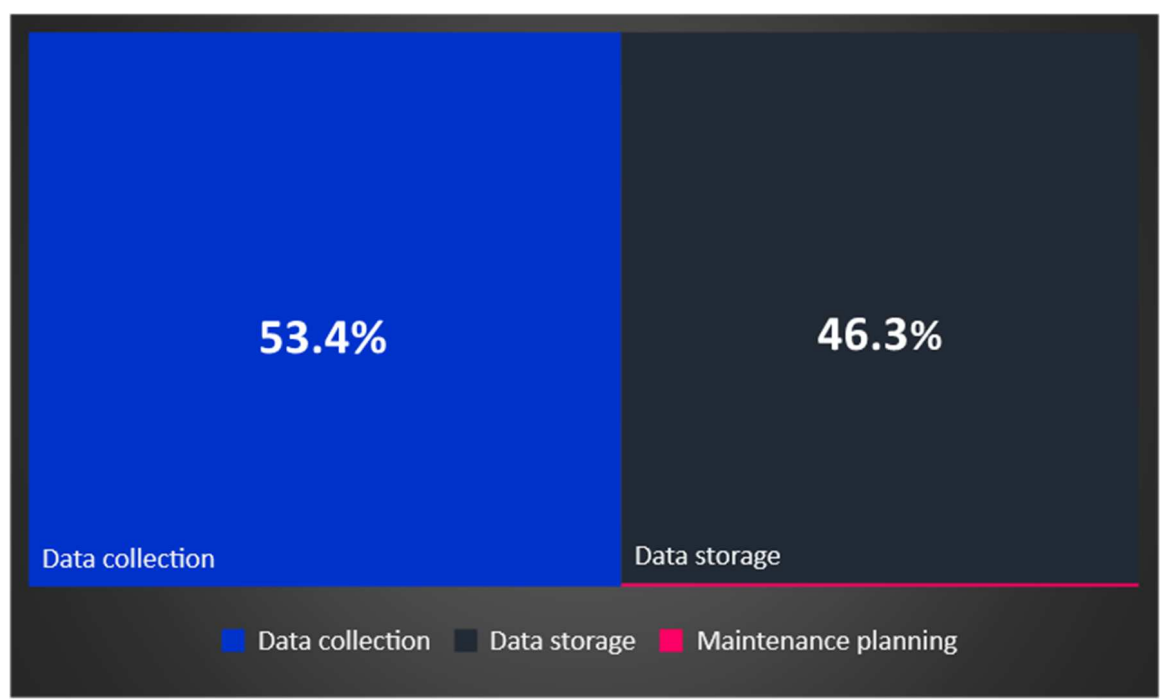

**Figure 6 The GWP distribution of data storage, data collection, and maintenance operation activities**

## 3.4  *Sensitivity analyses*

In the base case, the subgrade is considered to evolve from PF2qs to respectively PF2 and PF3 over 25% of the section's length each. While this assumption has been proposed based on a road constructor's expertise, no national statistics can be used to validate its national representativeness. The sensitivity of the results to this evolution is tested based on a "restricted range" (Figure 7) and an "extreme range" (Figure 8) of subgrade class evolutions. The second sensitivity analysis can be considered as extreme or rather unlikely, as up to the entire length of the subgrade is tested to be either upgraded or downgraded.

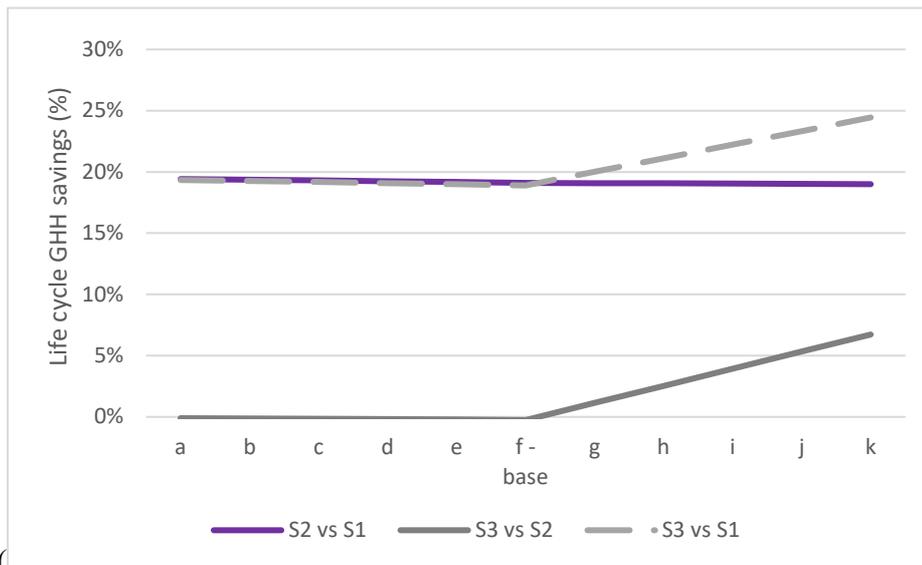

The two figures show that, if the subgrade's bearing capacity deteriorates over time (scenarios a to e), the structural reinforcement of the complete section classically carried out to respect the pavement design reliability is globally environmentally equivalent to minimal maintenance consisting of crack filling to delay alligator cracking, or rehabilitating short pavement sections with premature damage due to the worst-performing materials over the section. In this case, the advantage of using a PMS to tailor maintenance overtime may be masked by the safety margins taken in the design process due to the spatial heterogeneity of the pavement's behavior. On the contrary, if the bearing capacity of the subgrade improves, which may occur with the compaction of the pavement under heavy traffic (scenarios g to k), the results highlight that it is environmentally beneficial to know the condition of the pavement to adjust the maintenance with structural reinforcement. The PMS shows all its potential in these scenarios, as it allows to delay maintenance operations where they are not needed. Quantitatively, results show that savings of up to resp. 7 and 25% of the pavement life cycle GHG emissions can be reached using BIM to tailor maintenance (scenario S3) compared to resp. classic progressive or massive design and maintenance approaches (scenaris S1 and S2) on a restricted range of subgrade bearing capacity evolutions (Figure 7). When considering an extreme range of psubgrade bearing capacity evolutions, these gains reach up to resp. 14 and 30% over the pavement life cycle. If we restrain the scope to the service life and exclude the emissions from the original construction, these gains skyrocket to 47 and 65%.

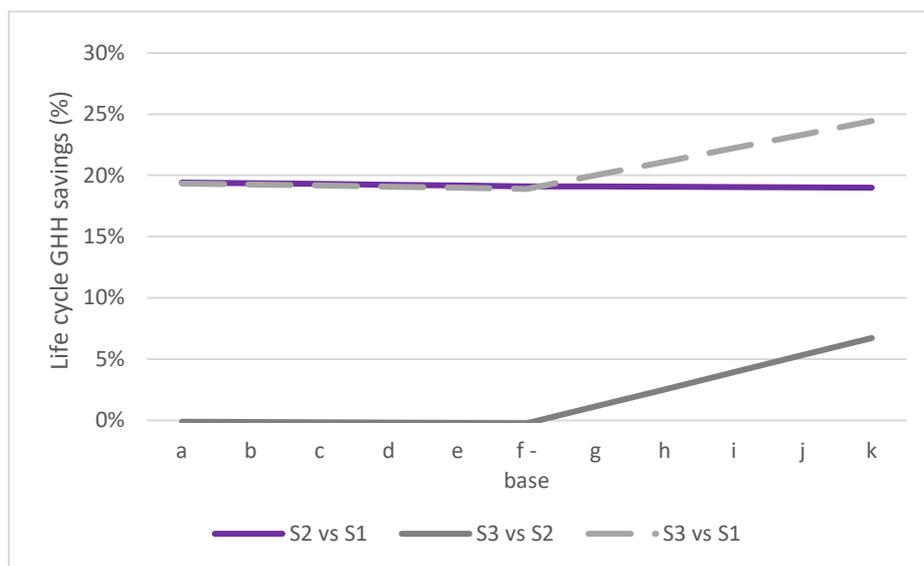

**Figure 7 GHG savings of the different pavement management scenarios over the life cycle "restricted range" sensitivity analysis**

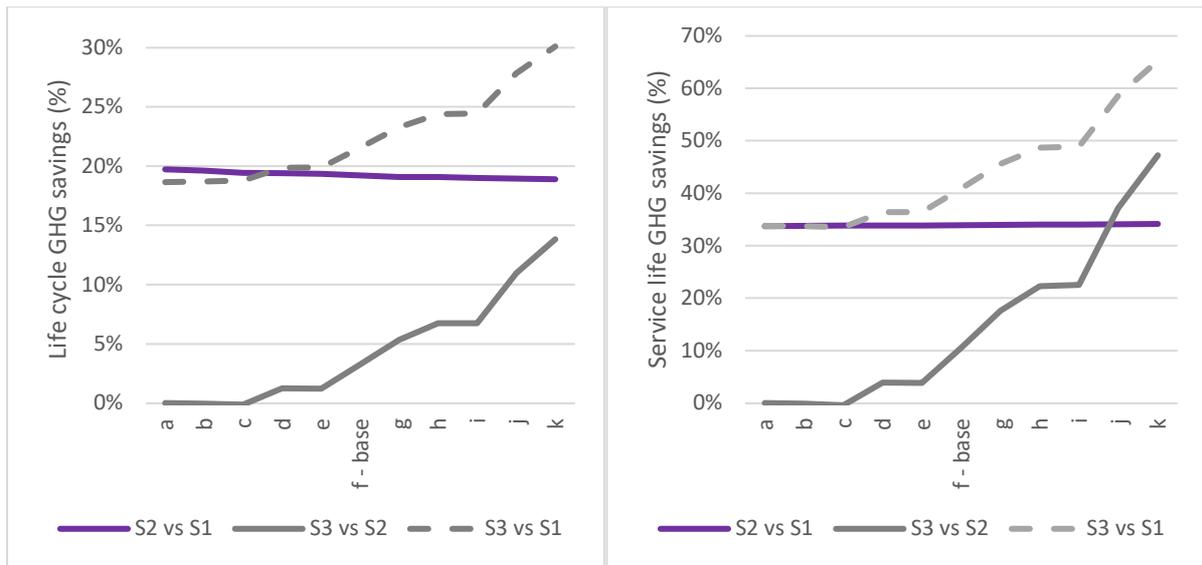

(a)                  (b)

**Figure 8 GHG savings of the different pavement management scenarios - over the complete life cycle (a) and on the service life (b) - depending on the evolution of the subgrade bearing capacity – "extreme range" sensitivity analysis**

# 4 Discussion

In this case study, we calculated the impact of using a PMS – the equivalent of a monofunctional I-BIM for pavement maintenance planning – on a high-traffic road life cycle carbon footprint. Nevertheless, I-BIM can provide 47 different types of technical support defined by Costin et al. (2018), and a multifunctional I-BIM would necessarily have different impacts. For instance, we estimated that an equivalent amount of GHGs are emitted over the pavement lifespan with a progressively designed pavement that is maintained with the support of a PMS when the bearing capacity of a subgrade deteriorated. But these results may only reflect French maintenance practices that could be too strict from the point of view of the mechanical maintenance of the structure. Thus, I-BIM could be used to break free from established rules and optimize structures over time according to desired criteria, for example, carbon emissions minimization. Indeed, monitoring the condition of pavements in real-time through instrumentation and using these data in an I-BIM to simulate the structural evolution over time would greatly advance maintenance management compared to what current structural simulation software programs allow. The big data generated and stored on the condition of the pavements and their environment could seed the development of algorithms based on artificial intelligence (AI) allowing for better simulate the behavior of road structures. Indeed, road design methods have gone from an empirical approach to a combination of mechanical physics and empiricism ("ME" methods) in the face of the inability of empiricism to capture the unlimited heterogeneity of pavement behavior due to limited data. The road sector has then turned to ME methods to reduce the errors of the physical models thanks to calibration coefficients. But ME methods are still limited to forecast pavement evolutions, and one might think that behavior models generated thanks to machine learning - and especially artificial neural networks - would perform better. Yet, AI algorithms are known to potentially consume a massive amount of electricity, and particular electronic elements whose environmental impact could be heavy (Strubell et al., 2019). Future studies should look into these aspects to ensure the development of a climate-positive I-IBM.

Moreover, we have shown that, although the carbon footprint of the PMS itself is negligible compared to construction operations, half of its emissions are due to the collection of data and the other half to their storage. Instrumenting the roads with sensors rather than surveying them with internal combustion engine (ICE) vehicles could help reduce the impact of the "maintenance planning" function of an I-BIM, but the sensors would also generate impacts, despite these impacts need to be investigated further (Pirson and Bol, 2021). In addition, the data generated could be much more massive and its storage could thus generate substantial GHG emissions. Thus, it would be appropriate in the future to assess the carbon footprint of multifunctional I-BIMs and to study the risk associated with the profusion of data.

Finally, the impact of road construction has been proven to be very limited compared to its usage (de Bortoli et al., 2022a; de Bortoli and Agez, Under review; Wang et al., 2012), and I-BIM could help optimize the management of pavements to reduce their environmental impact over their entire life cycle, including through their surface condition and geometry that impact vehicle consumption and aging (de Bortoli et al., 2022a; Chatti and Zaabar, 2012; Wang et al., 2012; de Bortoli et al., 2022b).

In any case, a multitude of different I-BIMs could be developed. And according to Bellman's principle of optimality, the sum of the optima of subsystems is different from the optimum of the system (Bellman, 1952). Also, optimizing recursively subsystems - each I-BIM's subsystem corresponding to one specific function - will be necessary to get close to the systemic optimum of the I-BIM over time. Depending on how I-BIM is used and in which technologic conditions, its impact will vary. For example, the environmental impact of the elements of the technosphere such as electricity (Alderson et al., 2012; Wolfram et al., 2016), metals (Watari et al., 2021), electronic components, and else, varied and will keep on varying in time and space. Forward-looking assessments are uncertain but necessary to project the potential consequences of I-BIM on climate. Additionally, we only considered GWP in this study. However, burden-shifting should be monitored and controlled. According to the Information Communication Technology scope published by ADEME, freshwater eutrophication and metal resources' would be the first two key burdens shifted with digitalization (Bio Intelligence Service, 2011), and they should be assessed closely when it comes to planning the development of BIM.

# 5 Conclusions

This article evaluates for the first time the environmental impact of the use of a BIM for infrastructure. More specifically, the carbon footprint of two road designs and three maintenance management approaches are compared: massive design accompanied by surface maintenance (scenario 1), progressive design combined with structural maintenance without monitoring of the bearing capacity of the subgrade (scenario 2), and the same progressive design with optimized maintenance thanks to a PMS recording the deflection of the pavement, the PMS being equivalent to a monofunctional I-BIM. This case study shows that the direct carbon footprint of I-BIM is negligible compared to the carbon footprint of pavement construction and maintenance operations over 30 years. In addition, the indirect effect of PMS, i.e. its consequences on emissions from construction operations taking place during the life of the pavement, are rather positive. Nevertheless, I-BIM is still novel, transitioning, and thus knowledge is limited: I-BIM can become a drag as a climate asset, and further and continued research will be needed to ensure its development serves the transition to carbon neutrality.

**Acknowledgment:** The idea of this study was initiated by the National Federation of Public Works in France (FNTP), which wishes to develop knowledge on the environmental impact of the infrastructure sector, particularly on the impact of its digitization on the climate. The authors want to thank the participation of the Technical Department of Eurovia Management in the design of the study on the construction aspect, and specifically Ivan Drouadaine – director of technics and research - for the reflection on the design-construction-maintenance alternatives. The authors also thank the French highway concessionaire Autoroute du Sud de La France (ASF), and especially Albane Hagnere – infrastructure project lead - and Sylvain Guilloteau – I-BIM lead - for their support in collecting data relating to the use of PMS (data storage and use of software), as well as Cécile Giacobi - pavement asset manager - for her insights and data on pavement monitoring and the use of PMS software.

**Funding source and role:** Anne de Bortoli: Investigation; Conceptualization; Methodology; Software; Validation; Visualization; Formal analysis; Project administration; Supervision; Writing - original draft; Yacine Baouch: Investigation; Software; Formal analysis; Visualization; Writing - original draft. Mustapha Masdan: Software; Validation; Visualization; Formal analysis; Writing – review, and editing.

313131